\documentclass[aps,prb,twocolumn,showpacs,superscriptaddress,floatfix]{revtex4}
\usepackage{graphicx}
\usepackage{color} 

\begin{document}
\title{Local Density of the Bose Glass Phase}
 
\author{K. Hettiarachchilage} 
\affiliation{Department of Physics, The College of New Jersey, Ewing, New Jersey 08628, USA}
\affiliation{Department of Physics \& Astronomy, Louisiana State University, Baton Rouge, Louisiana 70803, USA}

\author{C. Moore} 
\affiliation{Department of Physics, Southern University and A\&M College,
Baton Rouge, Louisiana 70813, USA}
\affiliation{Department of Physics \& Astronomy, Louisiana State University, Baton Rouge, Louisiana 70803, USA}

\author{V. G. Rousseau}
\affiliation{Department of Physics, Loyola University New Orleans, New Orleans, Louisiana 70118, USA}

\affiliation{Department of Physics \& Astronomy, Louisiana State University, Baton Rouge, Louisiana 70803, USA}

\author{K.-M. Tam}
\affiliation{Department of Physics \& Astronomy, Louisiana State University, Baton Rouge, Louisiana 70803, USA}

\author{M. Jarrell}
\affiliation{Department of Physics \& Astronomy, Louisiana State University, Baton Rouge, Louisiana 70803, USA}
\affiliation{Center for Computation \& Technology, Louisiana State University, Baton Rouge, Louisiana 70803, USA}

\author{J. Moreno}
\affiliation{Department of Physics \& Astronomy, Louisiana State University, Baton Rouge, Louisiana 70803, USA}
\affiliation{Center for Computation \& Technology, Louisiana State University, Baton Rouge, Louisiana 70803, USA}
\date{\today}

\begin{abstract}
We study the Bose-Hubbard model in the presence of on-site disorder in the canonical 
ensemble and conclude that the local density of the Bose glass phase behaves differently at incommensurate
filling than it does at commensurate one. 
Scaling of the superfluid density at incommensurate filling of
$\rho=1.1$ and on-site interaction $U=80t$ predicts a superfluid-Bose glass transition at
disorder strength of $\Delta_c \approx 30t$.   At this  filling the 
local density distribution shows skew behavior with increasing disorder strength.
Multifractal analysis also suggests a multifractal
behavior resembling that of the Anderson localization. Percolation analysis points to a
phase transition of percolating non-integer filled sites around the same value of disorder. Our findings support the
scenario of percolating superfluid clusters enhancing Anderson localization  near the superfluid-Bose glass transition. On the other hand, the behavior of the commensurate filled system is rather different. Close to the tip of the Mott lobe 
($\rho=1, U=22t$) we find a Mott insulator-Bose glass transition at disorder strength of 
$\Delta_c \approx 16t$. An analysis of the local density distribution  shows
Gaussian like behavior for a wide range of disorders above and below the transition. 
The behaviors of the superfluid-Bose glass transition call for a thorough finite size scaling analysis of percolation and multifractality to understand the universality of the transition.

\end{abstract}

\pacs{05.30.Jp, 03.75.Hh, 64.70.Tg, 61.43.Bn, 05.70.Jk, 72.15.Rn}

\maketitle 
\section{Introduction}
The Bose-Hubbard model \cite{Gersch} was originally proposed to demonstrate the existence
of a macroscopically occupied state under a repulsive interaction. 
By introducing quenched disorder \cite{Gimarchi,Fisher} this model exhibits a complex phase 
diagram. Many theoretical investigations of disordered interacting bosonic models
followed~\cite{DFisher,Fisher,Freericks,Scalettar,Krauth,Kisker,Kisker2,Carusotto,Amico,
Semerjian,Anders,Rancon,Yokoyama,Pai,Dawson,Lacki,Hettiarachchilage,Zhang,Sheshadri,
Priyadarshee,Gurarie,Yao,Meier,Soyler,Zuniga,Thomson,Ng,Hitchcock,Singh,Oosten,
Capogrosso-Sansone,Lin,Niederle,Batrouni,Hu,Pollet} early experiments on $^4$He films
absorbed on porous media~\cite{Finotello,Angolet,Bishop,Crowell,Csathy}. More recently, 
due to advances on optical lattice experiments, the Bose-Hubbard model has also become 
relevant in the reign of atomic physics~\cite{Jaksch,Bloch,Greiner}. Indeed, it quickly
becomes the most important venue for the physical realization of the Bose-Hubbard model \cite{Greiner}. 

In the absence of disorder the Bose-Hubbard model is rather well
understood, but the physics of the disordered model has shown to be much
complicated. An outstanding controversial issue is related to the quantum phase transition 
at commensurate filling. Early studies suggested that a direct superfluid Mott insulator
transition was unlikely though not fundamentally impossible\cite{Fisher}. A third phase,
the compressible and gapless Bose glass, intervenes between the superfluid and Mott insulator.  
Recent arguments justified the existence of the Bose glass upon the destruction of the 
Mott insulator based on the appearance of rare but compressible superfluid clusters \cite{Pollet,Griffiths,Gurarie}.  
Observation of a superfluid-Bose glass
transition has been reported in recent cold atoms experiments\cite{Pasienski}.


While the phase diagram of the disordered Bose-Hubbard model has been extensively studied,
the nature of the Bose glass has not received that much attention. 
A real space renormalization group study has claimed that the local density is not self
averaging for the Bose glass phase \cite{Hegg}. It has further been proposed that replica 
symmetry is broken at higher than two dimensions \cite{Thomson}. There are reports which
suggest that the Bose glass phase can be understood as a system of non-percolating
superfluid clusters \cite{Niederle}. But, a recent Quantum Monte Carlo study on the
related hard core Bose model suggests that the transition is not due to percolation\cite{Pablo}. 

A simple physical interpretation of the Bose glass phase, borrowed from  
Anderson localization, is that the virtually free bosons in the presence of a
sufficiently strong disorder potential localize \cite{Weichman}. The wavefunction of 
the Anderson model has been studied in great detail in recent years
\cite{Rodriguez,Ujfalusi,Castellani,Wegner1979,Wegner1980,Yakubo,Lindinger}. A prominent feature of the
localized phase is the skew distribution of its local density \cite{Schubert}. More
interestingly, around the critical point between the metallic and the localized
phase the wavefunction exhibits multifractal behavior \cite{Rodriguez,Ujfalusi,Moore}. If
the Bose glass can be interpreted as an Anderson localized phase, a
natural question is whether some of those behaviors can be
rediscovered in the Bose-Hubbard model. 

In this paper, we focus on the nature of the Bose glass and its transition to the
superfluid phase. In particular, we investigate the behavior of the local density distribution
at a commensurate, $\rho=1$, and an incommensurate filling,
$\rho=1.1$. We find the local density distribution broadens as the disorder increases, but
there are substantial difference between the systems at and away from the commensurate
filling. We perform multifractal and percolation analyses and find rather strong evidences that multifractal behavior exists near the Bose glass superfluid transition, where the percolation clusters formed by the non-integer filled sites can also be observed.

The paper is organized as follows. In Section II we introduce the model and the
parameters for our study. In Section III we discuss the effects of disorder on the
incommensurate superfluid phase. In Section IV we present the effect of disorder on the 
commensurate Mott phase and highlight the difference in the local density distribution for
the two fillings. We conclude in Section V.
In the Appendix, we provide additional details of the percolation analysis.

\section{Model} \label{Sec:Model}
The Hamiltonian for the disorder Bose-Hubbard model on a two-dimensional square lattice takes the form:
\begin{eqnarray}
  \label{Eq:Hamiltonian} \nonumber \hat\mathcal H &=& -t\sum_{\langle i,
j\rangle}\Big(a_i^\dagger a_j^{\phantom\dagger}+H.c.\Big)\\
& & +\frac{U}{2}\sum_i \hat n_i \Big( \hat n_i-1\Big)+{\Delta\sum_i \epsilon_i \hat n_i},
\end{eqnarray}
where $a_i^\dagger$ ($a_i$)  is the creation (annihilation) operator of a soft-core boson at lattice site $i$ with number operator $n_i=a_i^\dagger a_i^{\phantom\dagger}$. The sum $\sum\limits_{\langle i,j\rangle}$ runs over all
distinct pairs of first neighboring sites $i$ and $j$, $t=1$  is the hopping integral between neighboring sites, $U$ is the strength of the on-site interaction, $\epsilon_i $ is a uniformly distributed random variable in the interval $[-\frac{1}{2},+\frac{1}{2}[$, and $\Delta$ is the disorder strength. The inverse temperature is set at $\beta=L$ unless otherwise stated. 

We perform a quantum Monte Carlo study of this model within the canonical ensemble 
using the  Stochastic Green Function algorithm~\cite{SGF,DirectedSGF} with global space-time updates~\cite{SpaceTime}. 
As only a rather small system size (256 lattice sites) can be studied, the choice of ensemble may affect the data. As we are particularly interested in the differences between commensurate and incommensurate fillings,
we use the canonical ensemble in which the number of particles is fixed during the entire sampling process.  Unlike most of the  Quantum Monte Carlo methods the 
Stochastic Green Function algorithm allows to set the canonical ensemble rather easily~\cite{SGF,DirectedSGF,SpaceTime}.

\section{Introducing Disorder into the Superfluid Phase} \label{Sec:Densityn1.1}
We consider a system with incommensurate filling factor or average density $\rho=1.1$, which in the absence of disorder shows superfluid behavior. Then, we introduce  disorder and identify the critical point of the transition to a disordered phase. 
Our choice of the value $\rho=1.1$ does not have any intrinsic physical meaning. We expect similar results for other values close-by. However, 
we do not attempt to choose a value too close to $\rho=1$ due to 
the anticipated difficulties to locate the superfluid-Bose glass transition point numerically. 

\subsection{Superfluid Density}
We follow the standard procedure to detect a transition between  superfluid and  non-superfluid phases by monitoring the superfluid density, $\rho_{s}$. The Hamiltonian (\ref{Eq:Hamiltonian}) satisfies the conditions needed for using the conventional formula which
relates the winding number to the superfluid density ~\cite{Superfluid}. 
Then, the superfluid density, $\rho_s$, can be calculated via the winding number, $W$, as $\displaystyle \rho_{s}=\frac{\langle W^2\rangle}{4t\beta}$ where $\beta$ is the inverse temperature~\cite{Pollock}. 

\begin{figure}[!htb]
 \centerline{\includegraphics[width=0.5\textwidth]{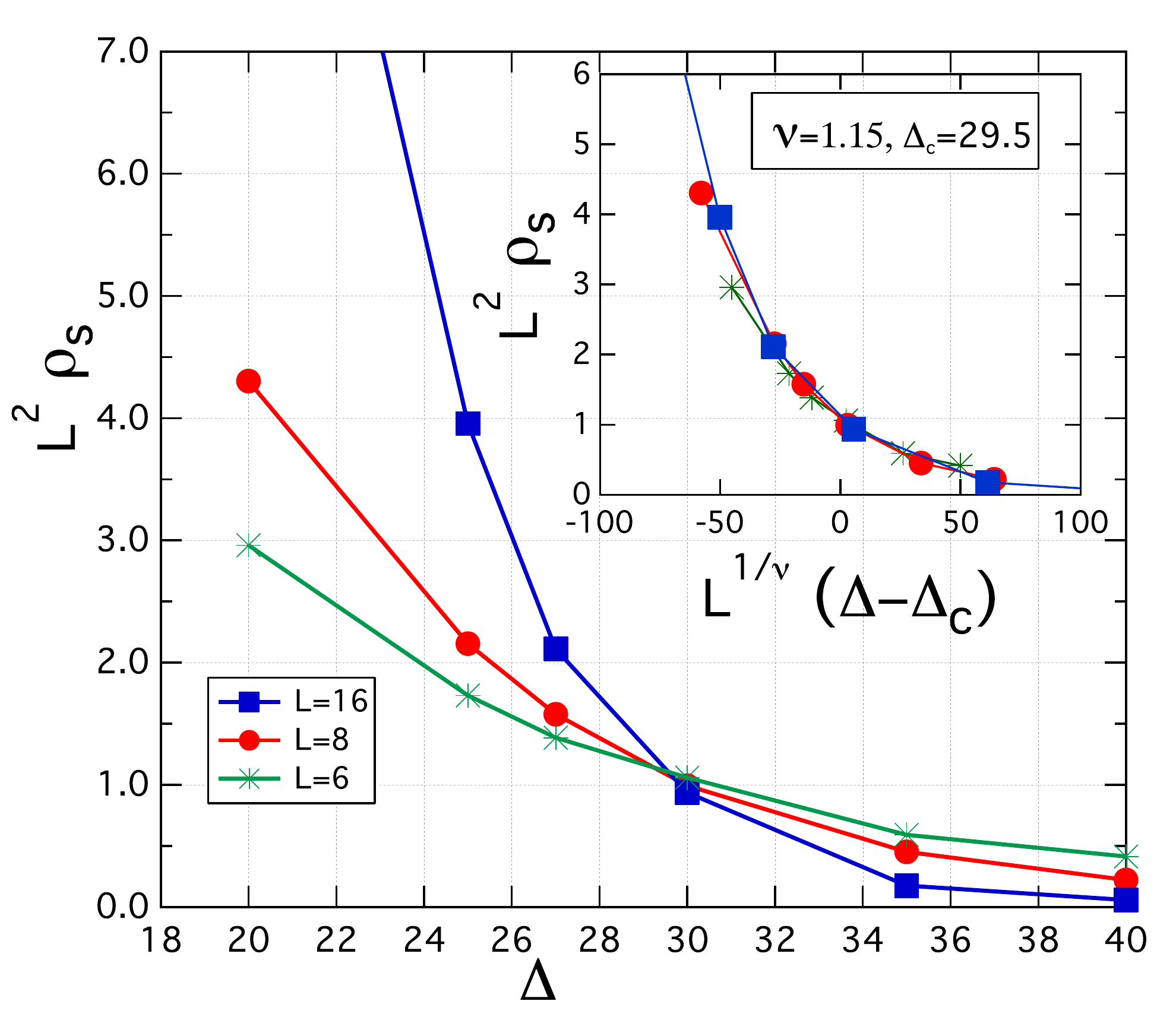}}
 \caption
    {(Color online) $L^2 \rho_s$ versus disorder strength, $\Delta$, for different system 
    sizes ($L=6,8,16$), density $\rho=1.1$, and on-site interaction $U=80t$. 
    The scaling analysis shows that the three curves cross at the critical disorder strength, $\Delta_c\approx 30t$. The data points are based on averaging the data from simulations of 1,000 disorder realizations.
    }
\label{Fig:Densityn1.1U80crossing}
\end{figure}

Fig. \ref{Fig:Densityn1.1U80crossing} displays $\rho_{s} L^2$ as a function of disorder strength $\Delta$ for three different system sizes: $L=6, 8,$ and $16$. 
In the neighborhood of the critical disorder strength $\Delta_c$, the superfluid density 
follows the scaling ansatz $\rho_s \sim L^{-z}g(L^{\frac{1}{\nu}}(\Delta-\Delta_c))$, 
where $z$ is the dynamical critical exponent, $\nu$ the correlation length exponent, and $g(...)$ a universal scaling function~\cite{Wang}. We based our
finite size scaling on the assumption that  $z=2$, the spatial dimension~\cite{Fisher}.
We locate the critical disorder at $\Delta_{c}=29.5t$ and the correlation length exponent $\nu=1.15$.

Our intent is not to pinpoint the critical point and its associated exponents with a very high precision, but to roughly locate the critical disorder and analyze the local density distribution for disorder strength close to the critical value. High precision calculations of the critical exponents of related models have been attempted in recent studies~\cite{Ng,Meier,Pablo,Wang}.  
For a more precise analysis one has to consider the scaling correction, and the goodness of fit, which could be rather challenging for the Bose-Hubbard model \cite{Ng,Pablo,Ujfalusi,Rodriguez,Moore,Wang}. We note that the value of $\nu=1.15$  we obtain is close to the latest estimates \cite{Ng,Pablo,Meier}. 

\subsection{Local Density Distribution}

\begin{figure}[!htbp]
  \centerline{\includegraphics[width=0.45\textwidth]{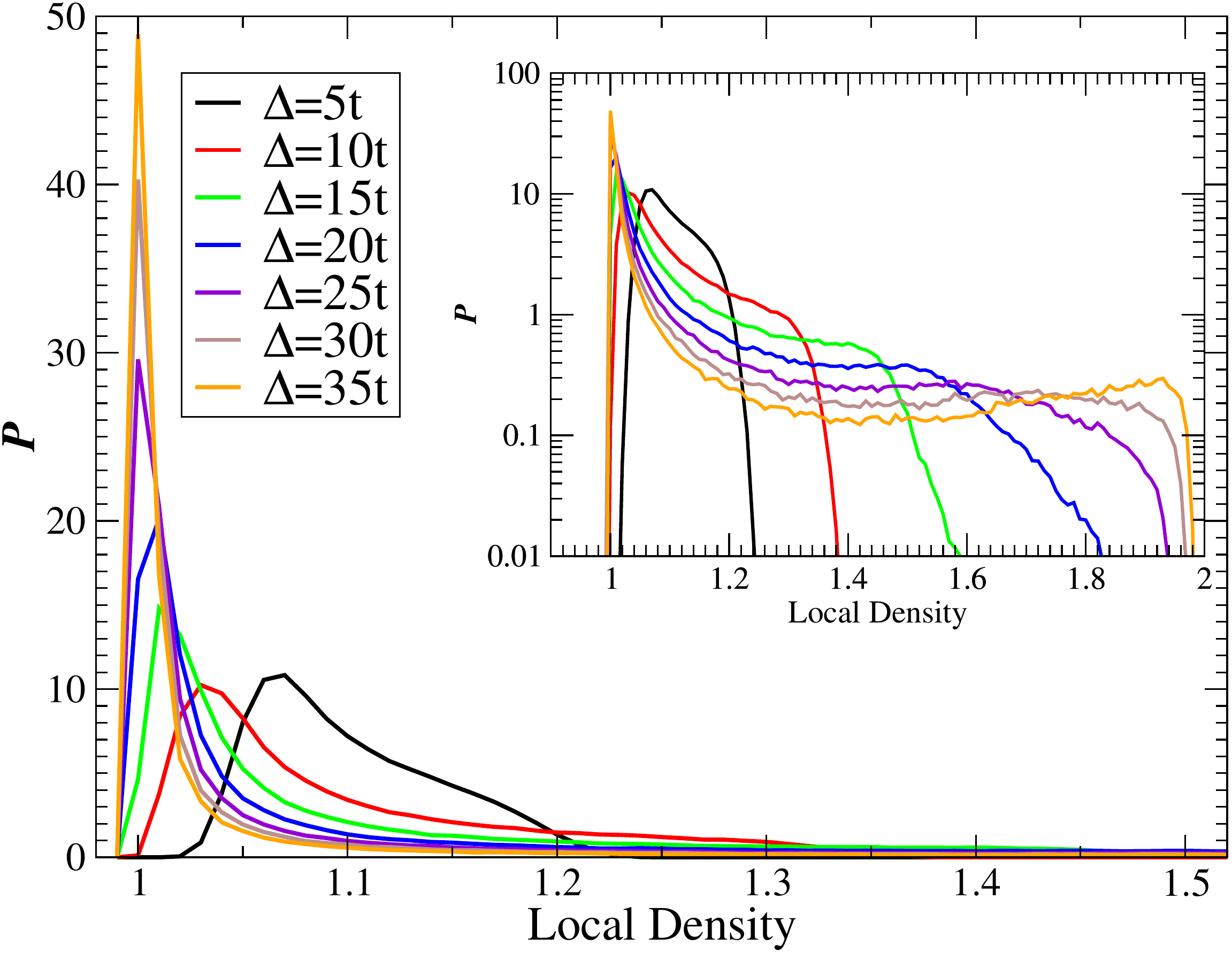}}
 \caption
    {(Color online) 
Histograms of the probability, $P$,  versus local density for system size $L=16$, density $\rho=1.1$, interaction
$U=80t$, and disorder strength between $\Delta=5t$ and $\Delta=35t$. Each calculation
includes 1,000 disorder realizations. The inset shows the same quantities in semi-log scale.
    }
\label{Fig:Densityn1.1U80}
\end{figure}

After we established the critical strength from scaling the superfluid density, we focus on the local density. 
Fig.~\ref{Fig:Densityn1.1U80} displays local density histograms 
for system size $L=16$, density $\rho=1.1$, interaction $U=80t$ for several disorder strengths. Each
calculation includes  1,000 disorder realizations. The inset shows the same 
quantities in a semi-logarithmic scale.

Fig. \ref{Fig:Densityn1.1U80} shows that while the behavior of the local density distribution 
is close to Gaussian at small disorder, it is already visibly deviates from Gaussian at the $\Delta=5t$, 
it becomes skew with a typical value very close 
to $\rho=1$ and a long tail, cut off around 2.0, for large values of the disorder strength. For

The skewness and long tail of the density distribution is the hallmark of the localized phase in the single particle Anderson model~\cite{MacKinnon,Schubert,Moore,Anderson}.
However, a true long tail distribution with no upper bound does not exist in the 
present model, as the local density is always cutoff at integer filling, most probably
due to the Hubbard energy penalty. We emphasize that the model 
we study is the standard Bose-Hubbard model without hardcore constraint.
Therefore these finding suggest that even in the Bose  glass phase the long tailed distribution does not extent all the way to infinity, but it is truncated due to the 
energy penalty for multiple occupation of a local site.

\begin{figure}[!htbp]
 \centerline{\includegraphics[width=0.45\textwidth]{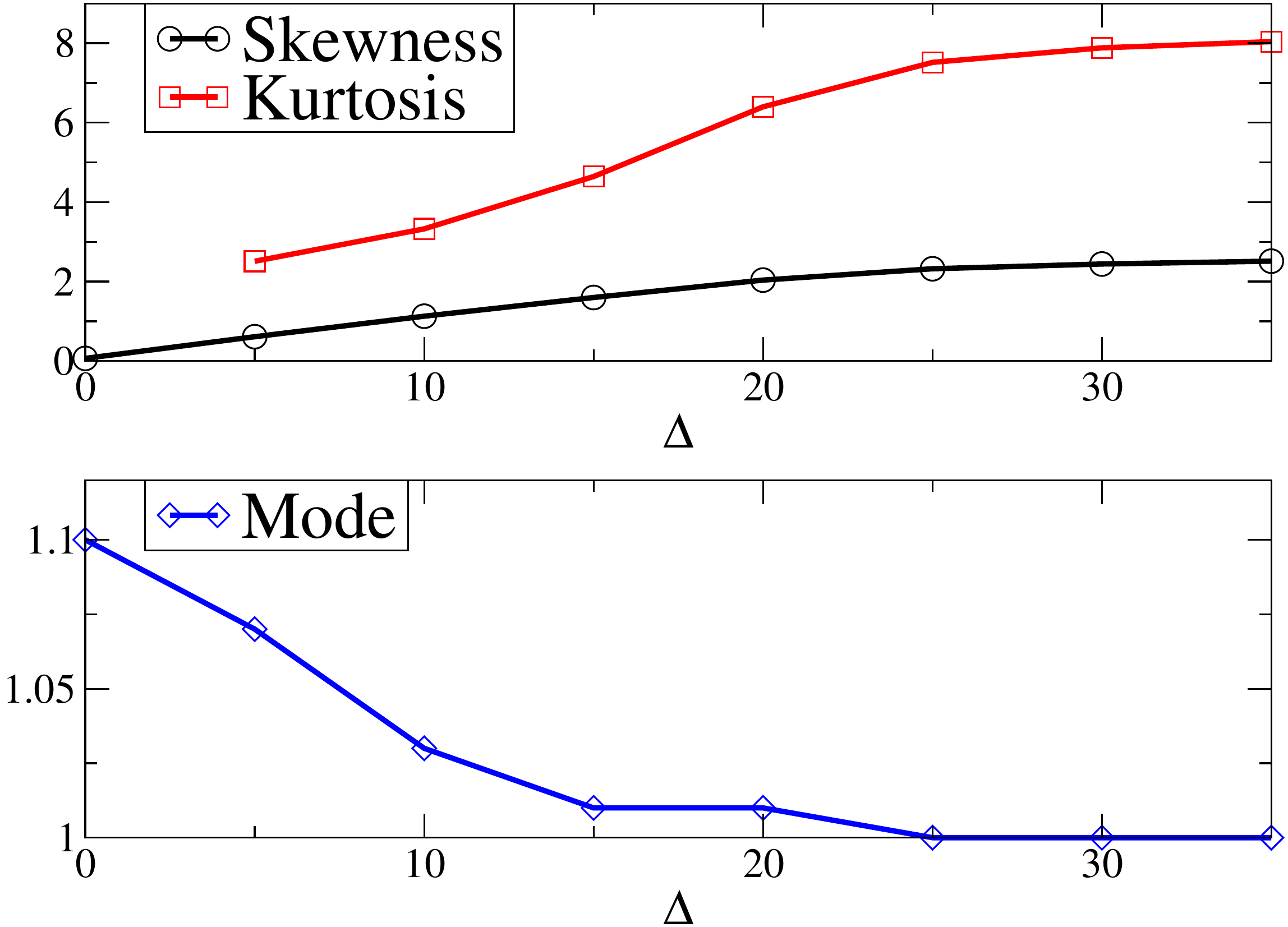}}
 \caption
    {(Color online) 
The skewness and kurtosis (upper panel), and mode (lower panel) of the local density
distribution as a function of disorder strength for system size $L=16$, density $\rho=1.1$, and interaction $U=80t$ as a function of disorder $\Delta$. The distribution for $\Delta=0$ is very narrow 
and its kurtosis cannot be calculated with enough precision. 
Mode is estimated from the histogram of the local density distribution with bin size $0.01$.}
\label{Fig:Skew_1.1}
\end{figure}

We corroborate these observations by calculating the skewness, kurtosis and mode of the
local density distribution as function of disorder strength. 
These measurements quantify the broadening of the distribution as the disorder increases. 
Fig.~\ref{Fig:Skew_1.1} shows that both the skewness and kurtosis grow with disorder 
strength to reach an apparent plateau for large disorder values.
The local density distribution for large disorder has kurtosis close to $8$ which is far from that the kurtosis of $3$ of a Gaussian. According to the typical medium theory for Anderson localization, the localized phase is signal by a typical local density equal to zero \cite{Ekuma,Dobrosavljevic,Janssen},  
we also plot the mode of the distributions in Fig. \ref{Fig:Skew_1.1}, bottom panel. 
We clearly see the mode of the distribution shifting from $1.1$ to $1$ as the disorder increases and settling at $1$ for disorder $\Delta$ larger than  $20t$. 

\subsection{Multifractal Analysis}

For $\rho=1.1$ the Bose glass can be considered as a diluted particles phase on a 
Mott insulating background where, in first approximation, the bosons exceeding  
integer occupation behave as independent particles in a random potential where
each local site is already occupied by one particle. The  quasiparticles in a two-dimensional random potential lattice localize unconditionally for the AI class~\cite{Anderson,MacKinnon,Atland,Evers}. 
Since Figs.~\ref{Fig:Densityn1.1U80} and \ref{Fig:Skew_1.1} support this point of view, 
we perform a multifractal analysis to look for similarities with the 
Anderson model \cite{Wegner1979,Wegner1980,Castellani,Rodriguez,Ujfalusi,Moore}.  

The multifractal analysis is based on the basic idea that the moments of a 
distribution cannot be described by a single exponent, but they are 
a continuous function of the order of the moment.
Calculations are performed  by dividing the system into different box sizes and 
calculating the moment for each box size. The moment is defined as:
\begin{equation}
Z_{q}(l) = \sum_{i}^{N_{l}} (m_{i}(l))^{q},
\end{equation}
where $m_{i}(l)$ is the local quantity (mass by convention) for the $i^{th}$ box, 
$N_{l}$ is the total number of boxes of linear size $l$, and $q$ is any 
real number. For our data with system size $L=16$, we choose $l=L/2,L/4$, and $L/8$. 
The multifractal dimension can be defined as the limit of the ratio of the logarithm 
of the moment to the logarithmic of the box size divide by $(q-1)$,
\begin{equation}
D_{q} = \frac{1}{q-1} \lim_{l\to0} \frac{\log(Z_{q}(l))}{\log l}.
\end{equation}
In practice, the limit of $l\to 0$ is estimated by linear extrapolation of $\log(Z_{q}(l))$ vs $\log l$.

\begin{figure}[!htbp]
 \centerline{\includegraphics[width=0.45\textwidth]{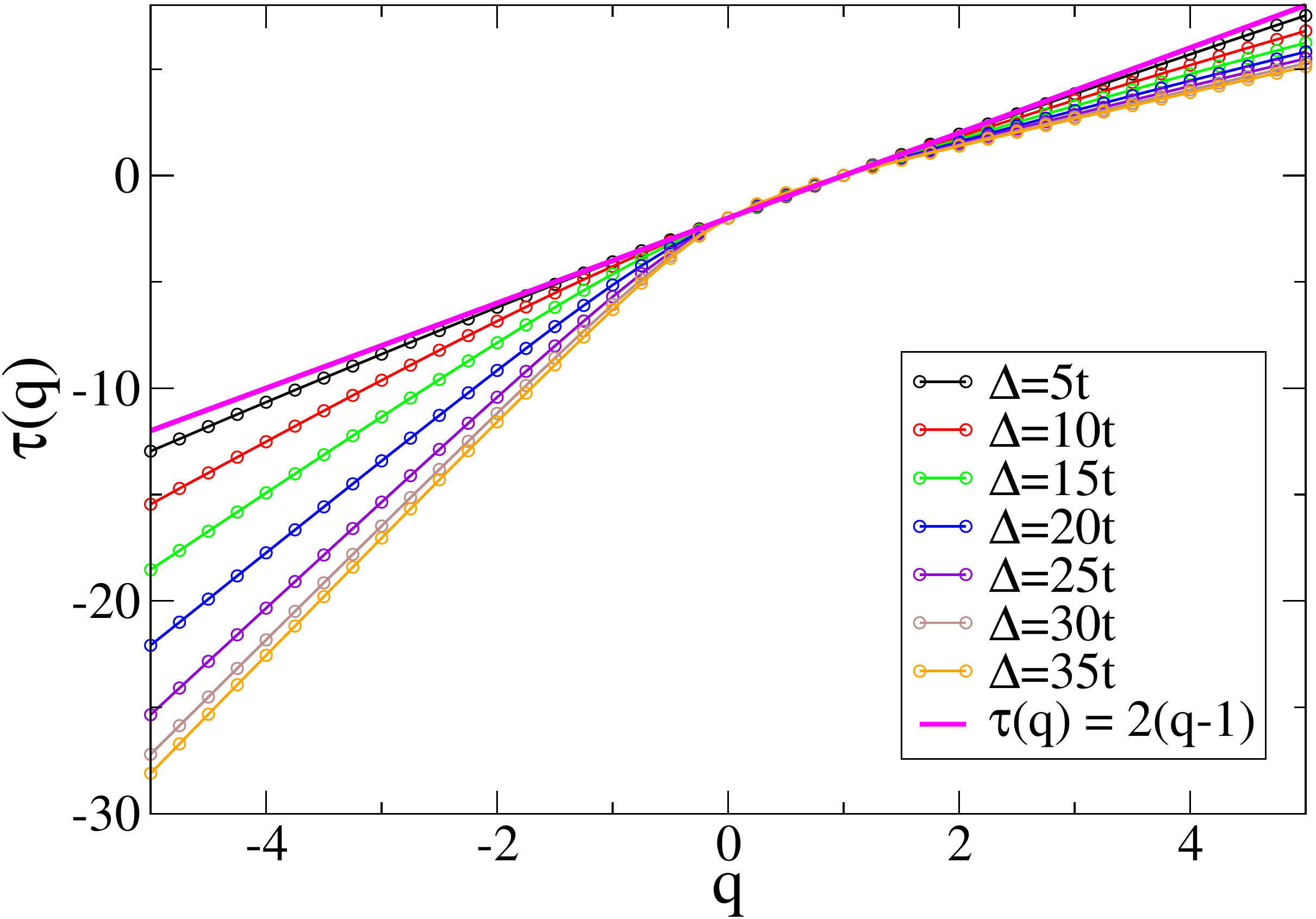}}
 \caption
    {(Color online) 
      Mass exponents of local density
      averaged over $1000$ disorder realizations. Local density measured for system size $L=16$,
      density $\rho=1.1$, and interaction $U=80t$ for disorder strengths $\Delta=5t, 10t, 15t, 20t, 25t,
      30t,$ and $ 35t$. The mass exponent of a non-fractal system is included 
      ($\tau(q) = 2 (q-1)$) for comparison.
    }
\label{Fig:Mass_exponents}
\end{figure}

One can also define the mass exponent,
\begin{equation}
\tau(q)=(q-1)D_{q}.
\end{equation}
There are two special points in the mass exponents: $q=1$ and $q=0$. For $q=1$, 
the mass exponent is always equal to zero provided that the input $m_{i}(L)$ is normalized.
For $q=0$, the mass exponent is equal to the negative of the dimension of the support, in 
this case the support is a square lattice, therefore $\tau(q=0)=-2$. A 
multifractal distribution is defined as a distribution which possesses a nonlinear 
dependence between the mass exponent $\tau$ and the order of the moment $q$\cite{Nakayama,Stanley,Salat}. For non-fractal systems,  their mass exponent is simply given as $\tau(q) = 2 (q-1)$ for a system with support on a square lattice. 

For the Bose-Hubbard model at incommensurate filling, we choose the mass as the deviation of the local 
density from an integer value: $m_{i}(L)=|\rho_{i}-1|$. This quantity is normalized for
each disorder realization before performing the multifractal analysis. Then $\tau(q)$ is
calculated for each realization separately, and averaged over 1,000 realizations for each
disorder strength, $\Delta$. Three different box sizes are used, $l=8,4$ and $2$, and $41$ different moments between  $q=-5$ and $q=5$. We use the package mfSBA for the
analysis~\cite{Saravia_2012,Saravia_2014}. Fig. \ref{Fig:Mass_exponents} displays the mass
exponent for different disorder strengths between $5t$ and $35t$. For system which does not
exhibit multifractality, the mass exponent is a linear function, $\tau(q) = 2 (q-1)$, also 
included in Fig. \ref{Fig:Mass_exponents}.
Note that for small values of the disorder $\tau(q)$ is very close to the non-fractal
limit. As the disorder increases the $\tau(q)$ curves bend further from the straight 
line, in particular for negative values of the moment. This is a typical signal of multifractality \cite{Chhabra,Halsey}. 

Another common measure of multifractality is the singularity
spectrum $f(\alpha)$. For each value of $q$, we can define the Hausdorff dimension as
\begin{equation}
f(q) = \lim_{l \to 0} \frac{1}{\log \ell} \sum_{i}^{N_{l}} M_{i}(l,q) \log M_{i}(l,q), 
\end{equation}
where $M_{i}(l,q) = (m_{i}(l))^{q} / \sum_{j}^{N_{l}} (m_{j}(l))^{q}$.
Similarly, for each value of $q$, we can define the average value of the singularity (distribution) strength as
\begin{equation}
\alpha(q) = \lim_{l \to 0} \frac{1}{\log \ell} \sum_{i}^{N_{l}} M_{i}(l,q) \log m_{i}(l).
\end{equation}

\begin{figure}[!htbp]
 \centerline{\includegraphics[width=0.45\textwidth]{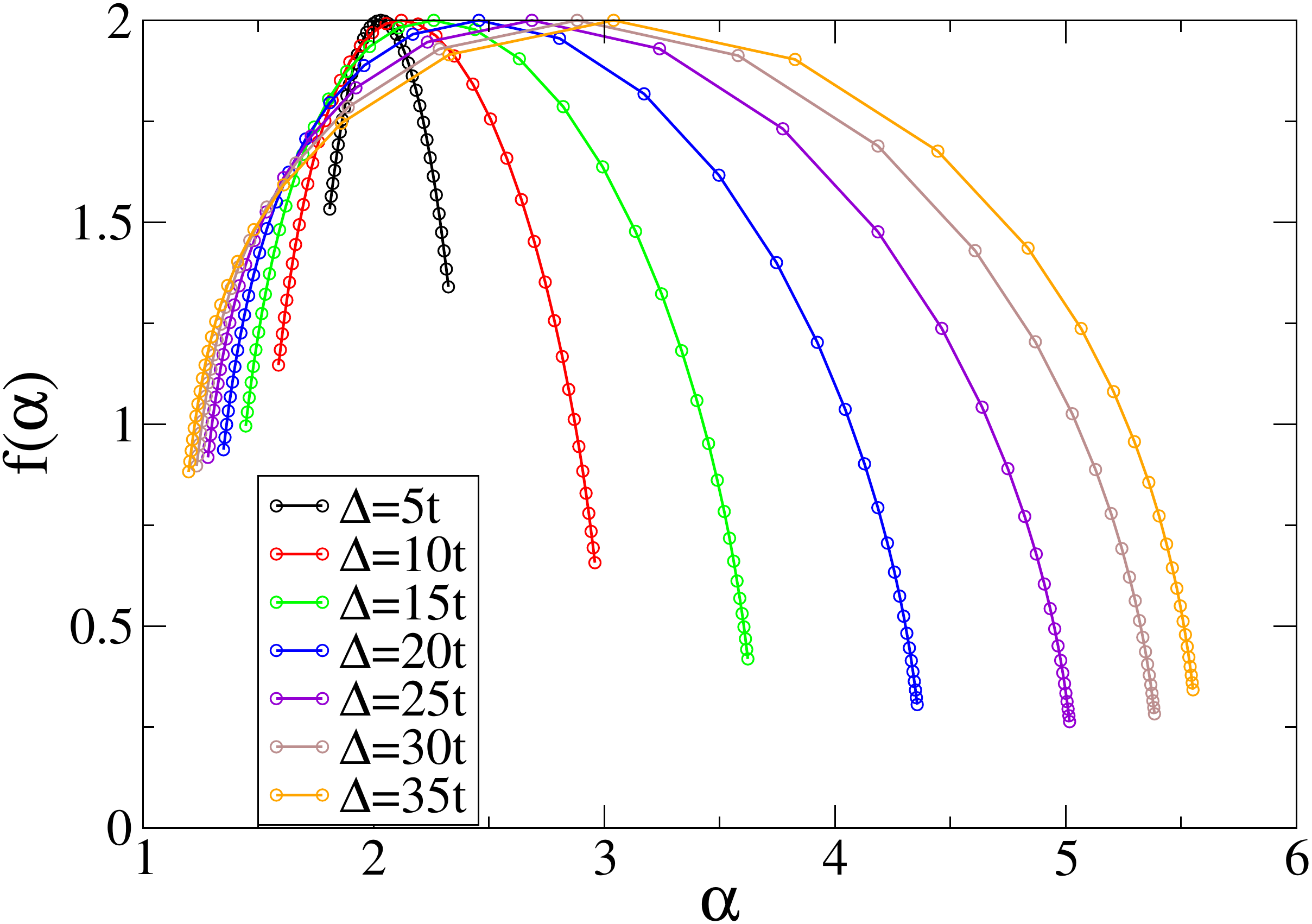}}
 \caption
    {(Color online) 
      Singularity spectrum of the local density
      averaged over 1000 disorder realizations. Local density measured for system size $L=16$,
      density $\rho=1.1$, and interaction $U=80t$ for the disorder strength values of $\Delta= 5t, 10t, 
      15t, 20t, 25t, 30t,$ and $35t$.
    }
\label{Fig:MultiFDensity1.1}
\end{figure}

The above equations set up an implicit relation between $f$ and $\alpha$ 
\cite{Chhabra,Halsey}. For systems which are non-fractal, the singularity
spectrum is concentrated around the point $(d,d)$, where $d$ is the system
dimensionality, $d=2$ in our case. On the contrary, for monofractal or multifractal
systems, an inverted curve with maximum at $(\alpha(q=0),f(q=0))$ is obtained, where
$f(q=0)$ is the Hausdorff dimension of the support. Therefore for a square lattice $f(q=0)=2$
\cite{Halsey}. The width of the singularity spectrum is a
measure of the degree of multifractality. A monofractal distribution has a very narrow 
spectrum while a strongly multifractal quantity displays a wide singularity spectrum.

To calculate $f(\alpha)$ we use the same set of $q$ values we 
employ in the calculation of $\tau(q)$. Fig.~\ref{Fig:MultiFDensity1.1} displays $f(\alpha)$
for several disorder strengths. For weak disorder within the superfluid phase the singularity spectrum
shows a rather sharp peak close to $(2,2)$. As the disorder increases $\alpha(q=0)$ increases from around 2 to a value  close to 3 for the largest disorder we explore. At the same time the singularity spectrum  
widens with increasing disorder. 

\begin{figure}[!htbp]
 \centerline{\includegraphics[width=0.45\textwidth]{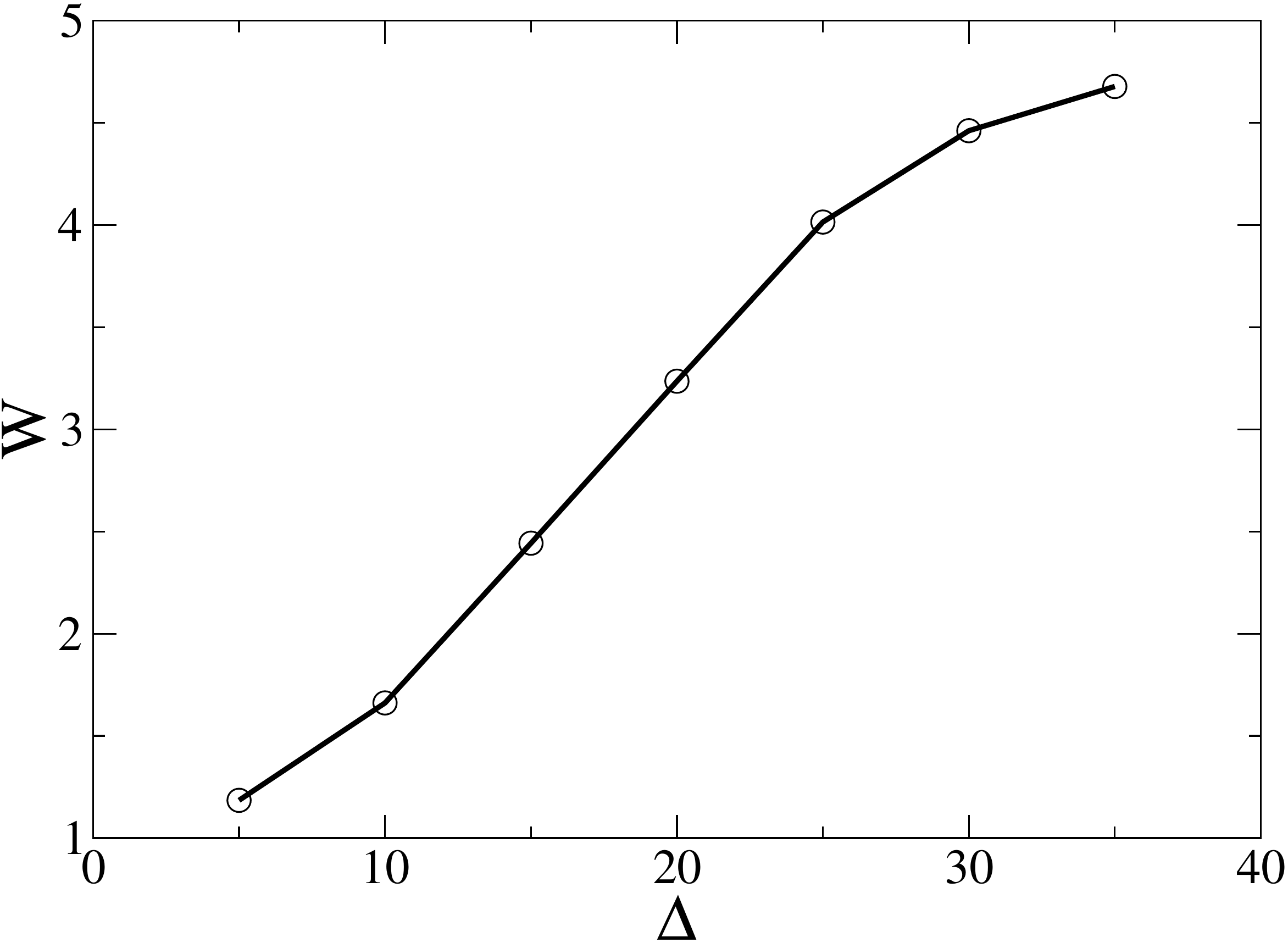}}
 \caption
    {(Color online) 
      Width ($W=\alpha_{max}-\alpha_{min}$) of the singularity spectrum of the local density
      averaged over 1000 disorder realizations for $L=16$, $\rho=1.1$, and $U=80t$. 
    }
\label{Fig:width1.1}
\end{figure}

We quantify the width of the distribution by fitting $f(\alpha)$ and then solving for the
two solutions when $f(\alpha)=0$ to obtain $\alpha_{min}$ and $\alpha_{max}$. The width
of the singularity spectrum can be defined as $W=\alpha_{max}-\alpha_{min}$~\cite{Stosic,Zhao}. 
Fig.~\ref{Fig:width1.1} displays $W$ as an increasing function of the disorder
strength. This widening increases faster between $\Delta=10t$ and $\Delta=25t$. For $\Delta > 25t$ the width of the 
spectrum still increases but at a smaller rate. 

Notice that the discussion and data presented in this section are not a multifractal finite size scaling analysis as has been done recently for the non-interacting models which possess Anderson localization transition
\cite{Rodriguez,Ujfalusi,Moore,Lindinger}. The $\tau$, $\alpha$, and $f$ are estimated by using Eq. (2) to (6) for a system of size $L=16$.
The notion of multifractality describes a system with scale invariant fluctuations which cannot be reduced to a single exponent. In general, scale invariance exists only at a second order transition point, which presumably is the superfluid-Bose glass transition within our model. For this very reason, one should only expect multifractality at exactly the critical value of disorder. The present analysis does not verify the scale invariance and it cannot pinpoint the value of the critical disorder based on multifractal finite size scaling analysis. Our data for the mass exponents and the singularity spectrum provide good evidence of multifractal behavior but it is not a definite proof.

\subsection{Percolation Analysis}

Since the early studies of the disordered Bose-Hubbard model, percolation has been
considered as a mechanism to understand the superfluid to Bose glass transition  \cite{Gao,Sheshadri_1995,Niederle,DellAnna,Buonsante,Barman}. However, 
there are some difficulties in using percolation as a criterion to identify the transition. First, the choice of the local physical quantity is important. In this study  we focus on the local density, but it is not entirely clear  whether it 
is unique or even a proper choice. Second, regardless the method, local mean field 
or quantum Monte Carlo, the precision of the measured local quantity is limited. 

\begin{figure}[!htb]
 \centerline{\includegraphics[width=0.45\textwidth]{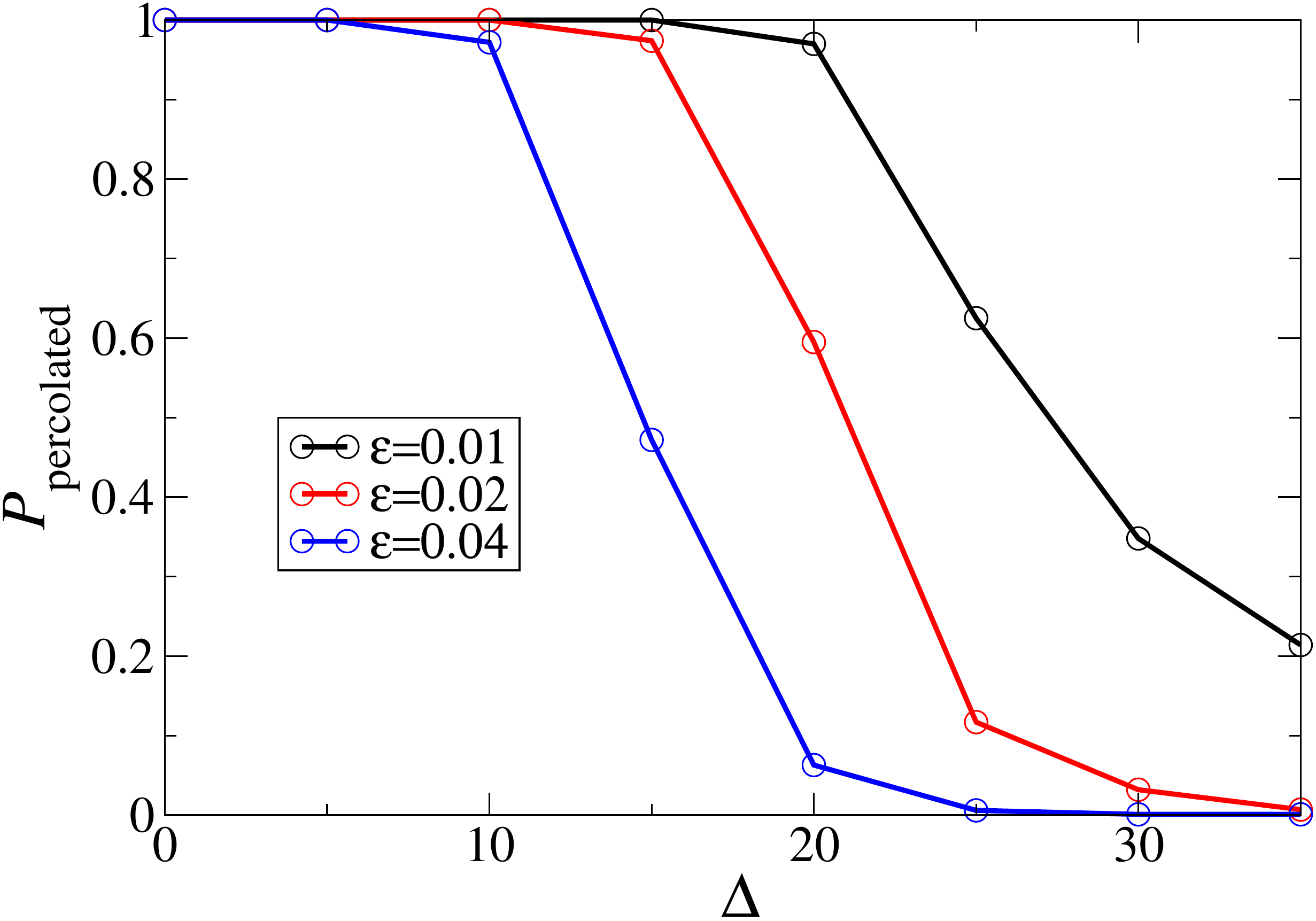}}
 \caption{(Color online) The probability of finding a percolating cluster of non-integer
 filling. Three cutoffs for integer filling are shown, $\epsilon=0.01,0.02,$ and $0.04$ 
 for the black, red, and blue lines, respectively. A local site is consider with
 integer occupation number if $|\rho_{i}-1| < \epsilon$. 
 We define the probability of a
 percolating realization as $P_{\text{percolated}} = N_{\text{percolated}}/N$.
$N_{\rm{percolated}}$ is the number of realizations with at
 least one percolated non-integer filling cluster. $N=1000$ independent realizations are used for each data point.
}
\label{percolation_probability}
\end{figure}
In our approach we need to choose a cutoff which discerns the sites with
integer local occupation number from those with non-integer occupation.
If a local site meets the criteria $|\rho_{i}-1| < \epsilon$, it is considered having
an integer occupation number. The cutoff is clearly influenced by the precision of the measured quantity. We thus choose three different cutoffs, $\epsilon=0.01,0.02,$ and $0.04$; 
where $\epsilon=0.01$ is a realistic estimate for the smallest cutoff. We do not attempt to choose a smaller cutoff, as it would be too close to the Monte Carlo sampling error. 
Since, the local density is not an averaged quantity over the lattice, 
its measurement is generally more prone to carry a large statistical error. 
Fig. \ref{percolation_probability} shows  the probability of a system with a non-integer
percolating cluster as a function of disorder for these three different cutoffs. 
$P_{\rm percolated}= N_{\rm percolated}/N$, where $N_{\rm percolated}$ is 
the number of realizations with at least one percolated cluster of non-integer filled sites out of a total of $N$ realizations. 
See the appendix for the definition of percolation and examples of randomly chosen
realizations for several values of disorder strength. 

With the cutoff $\epsilon=0.01$, the probability of a percolating cluster becomes $50\%$
for $\Delta \simeq 27t$, which is slightly smaller than the critical disorder strength of $29.5t$ 
we found scaling the superfluid density.  Most percolation transitions are second order,
therefore one can attempt to perform a finite size scaling to locate the critical point and
its exponents \cite{Kirkpatrick,Shante}. Given the available system sizes, we do not
attempt to perform a more detail finite size scaling. 

\section{Introducing Disorder into the Mott-insulating Phase} \label{Sec:Densityn1}
In the absence of disorder the Bose-Hubbard model at integer fillings is well understood. For strong interaction the ground state is a Mott insulator. According to previous studies, the Bose glass phase can appear from very weak disorder \cite{Soyler}. We note that a recent study suggests the Bose glass phase at weak disorder is anomalous \cite{Wang_MG}. We are mostly interested in the local density distribution near the Mott insulator to a gapless Bose glass. As we did for the case of $\rho=1.1$ in the previous section, we first established the critical value of disorder at a fixed interaction. Since the tip of the Mott insulator lobe occurs at $U_c \approx 16.7t$ \cite{Soyler}, we decide to introduce disorder at a slightly large value of $U=22t$. 

\begin{figure}[!htb]
 \centerline{\includegraphics[width=0.5\textwidth]{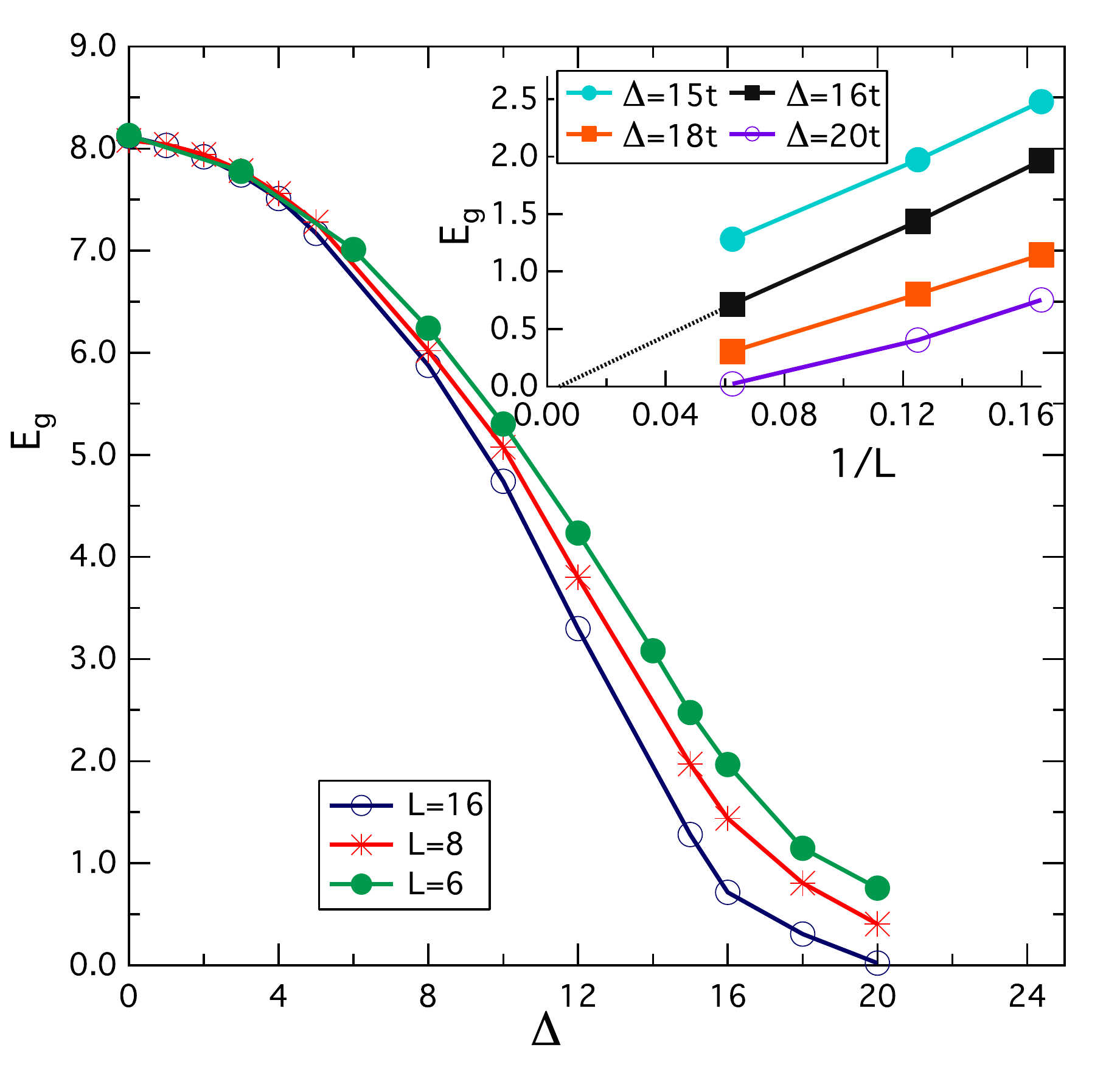}}
 \caption
    {(Color online) 
    Mott insulator gap $E_g$ versus disorder strength, $\Delta$, for different system
    sizes, $L=6,8,$ and $16$ at interaction $U=22t$. Data from $100$ disorder realizations are averaged for each data point. The inset displays $E_g$ 
    for several values of $\Delta$ as a function of $1/L$. By extending those
    curves, we find $\Delta_c \approx 16t$. The corresponding inverse temperatures for 
    the linear system sizes are $\beta t=12$ for $L=6$, $\beta t=16$ for $L=8$, and $\beta t=32$ for $L=16$. Data points are simulation results, lines are guides to the eye.
    }
\label{Fig:Densityn1U22gap}
\end{figure}


First, we look at the excitation gap. 
It has been suggested that there is no direct Mott insulator-superfluid transition \cite{Soyler,Gurarie,Pollet}, thus the vanishing of the particle excitation gap corresponds to the Mott insulator-Bose glass transition. The Mott gap is calculated as follow. We obtain the chemical potential by adding a particle to the system as, $\mu_1=E(N+1)-E(N)$ and also by removing a
particle from the system as $\mu_2=E(N)-E(N-1)$. The Mott gap is given as $E_g= \mu_1-\mu_2$. 
Fig.~\ref{Fig:Densityn1U22gap} displays the change of the energy gap, $E_g$, with 
increasing  values of $\Delta$ for three different system sizes, $L=6,8,16$ at $U=22t$ for 
$100$ disorder realizations. Since we are dealing with finite systems we find a finite gap for each $\Delta$ we consider and we need to do a finite size scaling to infer the value of the gap. The inset in Fig.~\ref{Fig:Densityn1U22gap} shows $E_g$ as a function of
$1/L$ for $\Delta=15t, 16t, 18t$ and $20t$. By extrapolating $E_g$ versus $1/L$ for different values of $\Delta$\, we extract a value 
of the critical disorder of $\Delta_c \approx 15.7t$.

\begin{figure}[!htbp]
 \centerline{\includegraphics[width=0.5\textwidth]{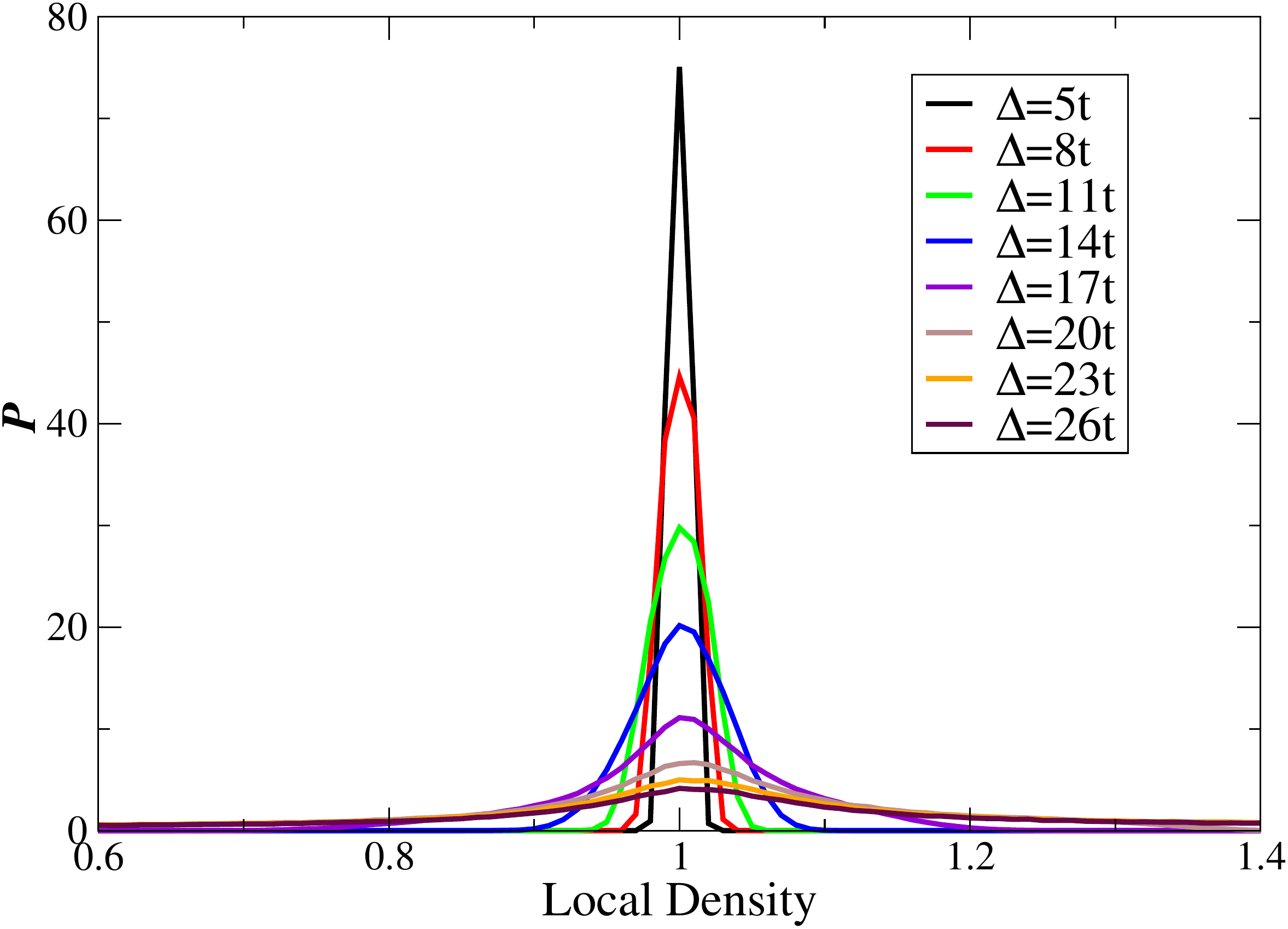}}
 \caption
    {(Color online)  
   Histograms of the probability, P, versus local density for disorder strength of $\Delta=5t, 8t, 11t, 14t, 17t, 20t, 23t$, and $26t$. The system size is $L=16$, the density is $\rho=1.0$, and the interaction is $U=22t$. $1600$ disorder realizations are calculated for each vale of disorder strength.
    }
\label{Fig:Densityn1U22}
\end{figure}

Fig.~\ref{Fig:Densityn1U22} displays the histogram of
the local density for $1600$ disorder realizations, system size $L=16$, $\rho=1.0$, 
$\beta t=16$, and $U=22t$ for disorder strength, $\Delta$, between $5t$ and $26t$.
The probability distribution of the local density for systems with weak disorder is
Gaussian like; the distribution does spread out with increasing disorder but, unlike the 
$\rho=1.1$ case, its skewness is small and the local density spreads over both sides 
of the peak. This remains the case even for large values of
disorder when the system is far from the Mott insulator phase ($\Delta_c \approx 15.7t$). 

\begin{figure}[!htbp]
 \centerline{\includegraphics[width=0.45\textwidth]{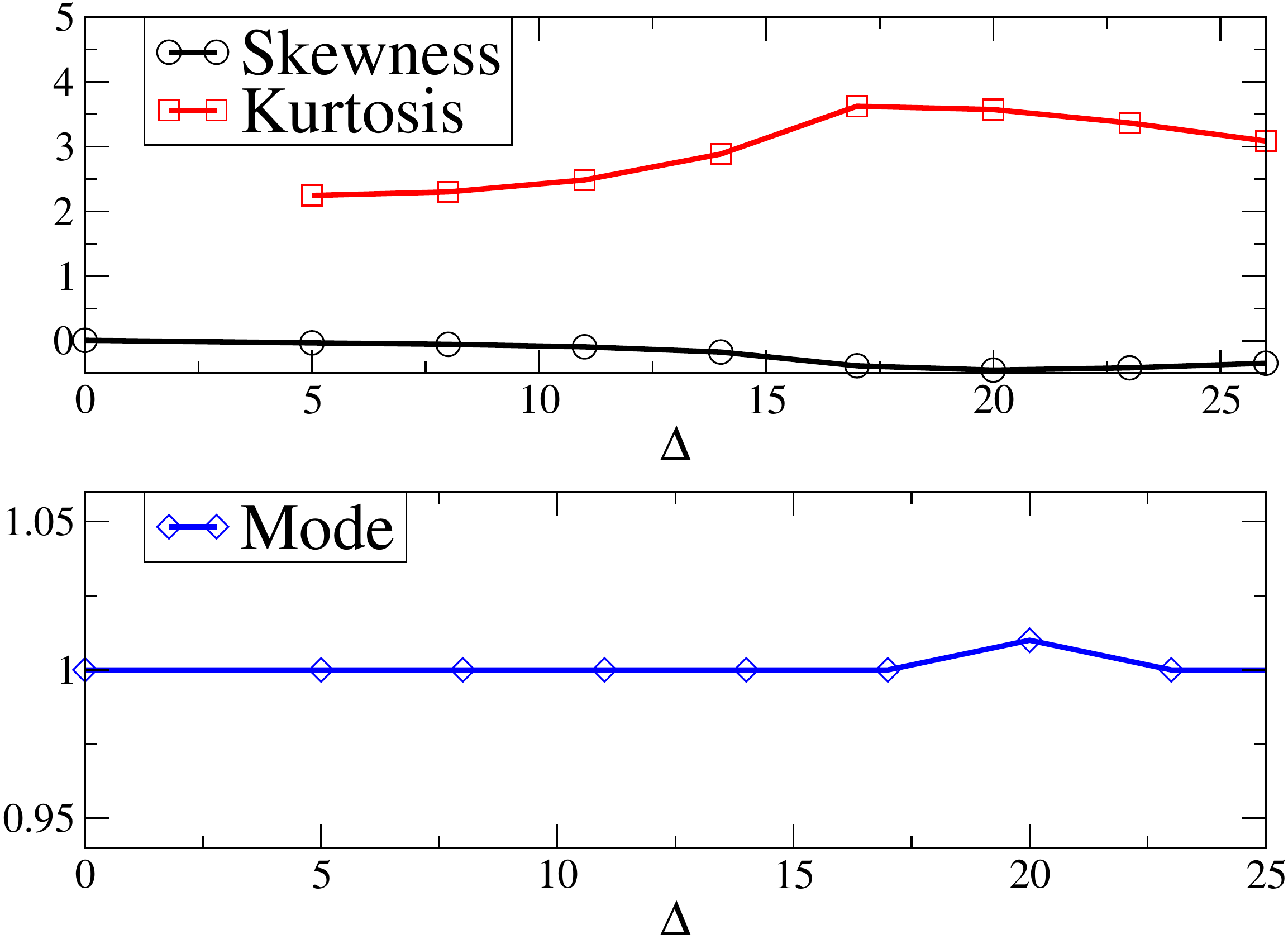}}
 \caption
    {(Color online) 
The skewness and kurtosis (upper panel), and mode (lower panel) of the local density
distribution as a function of disorder strength for system size $L=16$, density $\rho=1.0$, and interaction $U=22t$. 
The distribution for $\Delta=0$ is very narrow and its kurtosis cannot be calculated with
enough precision. Mode is estimated from the histogram of the local density distribution
with bin size $0.01$.}
\label{Fig:Skew_1.2}
\end{figure}

We further corroborate these observations by calculating the skewness, kurtosis, and mode
of the distribution as a function of disorder. Fig. \ref{Fig:Skew_1.2} shows those
quantities and confirms our findings. Both the skewness and kurtosis are greatly reduced compared with the values for $\rho=1.1$.  In particular the kurtosis, which can be
interpreted as a measure of the density of outliers, is fairly close to $3$ even for rather
strong disorder far away from the Mott insulator phase. In contrast with the case of
$\rho=1.1$ the mode is fixed at a constant value $\sim 1$.

We conclude that for $\rho=1$ the validity of the analogy with Anderson localization is
obscure since the picture of single particle in a disorder potential may not be valid. 
A many particle picture might be needed to explain the Bose glass phase at integer fillings.

\section{Conclusion} \label{Sec:Conclusion}

We study the spatial structure of the disordered Bose glass phase at both incommensurate
and commensurate filling. We analyze our results at 
incommensurate filling based on a simple picture of the single particle Anderson
localization. Given this picture, we test some of the characteristics of Anderson
localization, such as the skewness of the distribution and multifractality. We find that
for incommensurate filling ($\rho=1.1$), the local particle density has a skew
distribution, and the multifractal analysis shows resemblance to that of the single
particle Anderson localization. We also perform a percolation analysis to find that the 
probability of non-integer filling cluster does show a qualitative change near the
transition between Bose glass and superfluid. The difficulty in precisely  defining  
integer filling and the limitation in the available system size remain hindrances for a
definite answer. If the transition is simply a standard classical percolation transition,
then multifractality should not exist. A plausible scenario to reconcile multifractal and
percolation behavior is that almost percolating clusters enhance Anderson
localization. It is worthwhile to mention that the notion of percolation in the local
superfluid amplitude, $\psi_{i}=<a_{i}>$, enhancing the superfluid to Bose glass transition
due to localization has been proposed before \cite{Sheshadri_1995}. This picture does not
preclude multifractality due to localization at the critical point. On the other hand, the 
commensurate ($\rho=1$) case shows very different behavior. The skewness and the moment of
the local density distribution are greatly reduced  when compared with the values obtained
at incommensurate filling even when the system is far away from the Mott insulating phase. Clearly the local density distribution of the Bose glass at commensurate and incommensurate fillings cannot be described using the same picture. In particular, the single particle picture as that in the Anderson localization should fail. 


\section{Acknowledgment}
This work was supported by NSF OISE-0952300 (KH, VGR and JM). Additional support was
provided by the NSF EPSCoR Cooperative Agreement No. EPS-1003897 with additional
support from the Louisiana Board of Regents (CM, KMT and MJ). Additional support (MJ) was provided by NSF Materials Theory grant DMR1728457.
This work used the Extreme Science and Engineering Discovery Environment (XSEDE), 
which is supported by the National Science Foundation grant number ACI-1053575, and the high performance computational resources provided by the 
Louisiana Optical Network Initiative (http://www.loni.org). 

\appendix*

\section{Patterns of Percolating Clusters}

We discuss the percolation of non-integer filling clusters in Section III. In this appendix
we randomly pick 32 realizations from four disorder strengths ($\Delta=15t,25t,30t,35t$) to
illustrate the change of the number of percolating clusters as a function of disorder. The
cutoff criteria for a local site with integer filling is defined as 
$|\rho_{i}-1|< \epsilon$. Figures below are for $\epsilon=0.01$. Each realization contains
$16 \times 16$ sites.  The black and white squares represent sites with integer and 
non-integer occupation numbers, respectively. The blue area represents the cluster 
formed by the non-integer occupied sites. The cluster is defined starting at the top
and contains all the sites with non-integer occupation which are connected. 
The realization is considered as percolated if there is one non-integer filling cluster
which spans from the top to the bottom of the lattice. Since periodic boundary conditions
are used in the calculation, this definition may underestimate the value of the disorder
strength for the percolating cluster. For weak disorder, deep in the superfluid phase
$\Delta=15t$, all the realizations are percolated. As the disorder increases, more and 
more realizations break into isolated fragments of non-integer filling sites.

\begin{figure}[!htbp]
 \centerline{\includegraphics[trim={0 2cm 0 0}, width=0.5\textwidth]{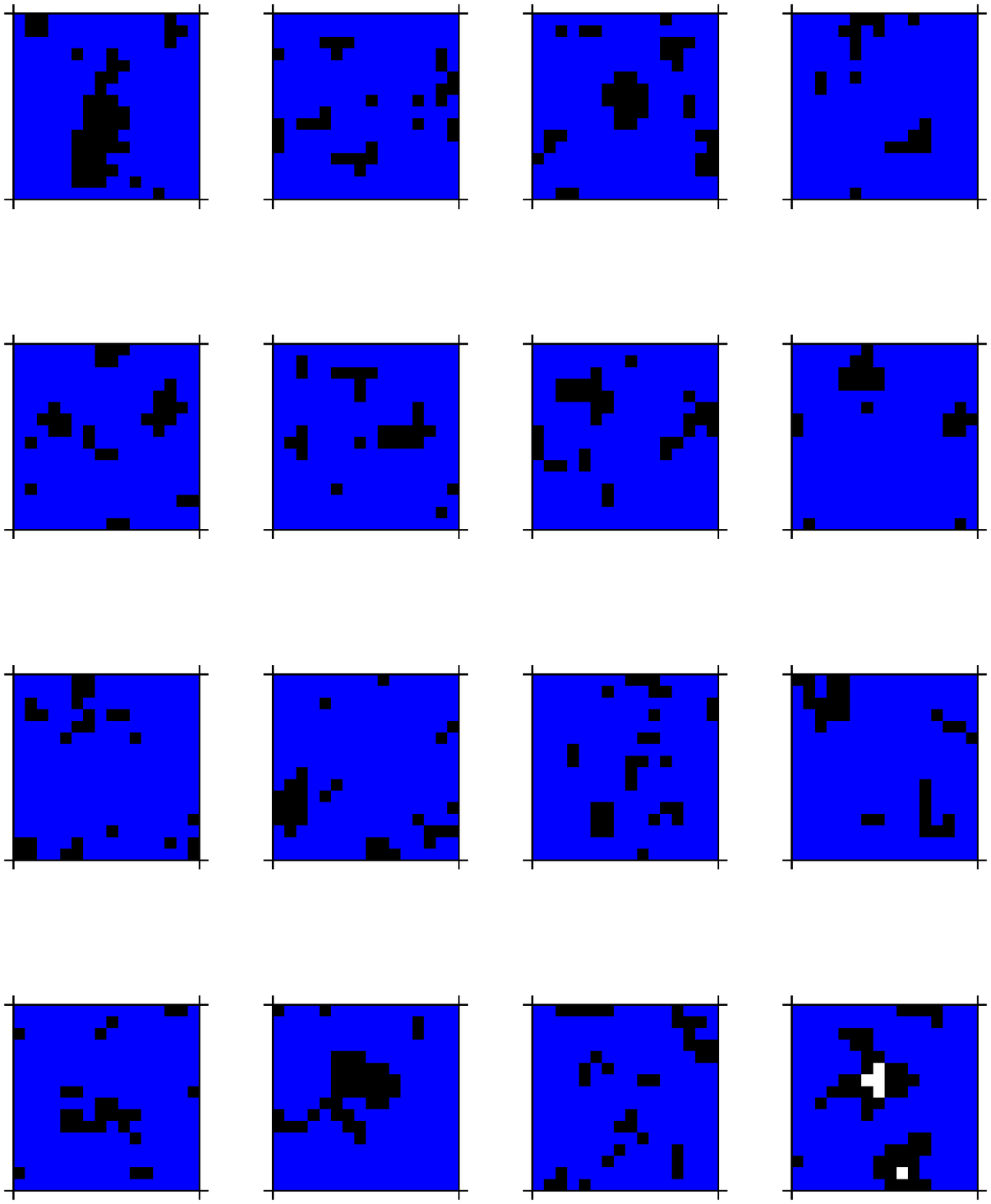}}
  \centerline{\includegraphics[width=0.5\textwidth]{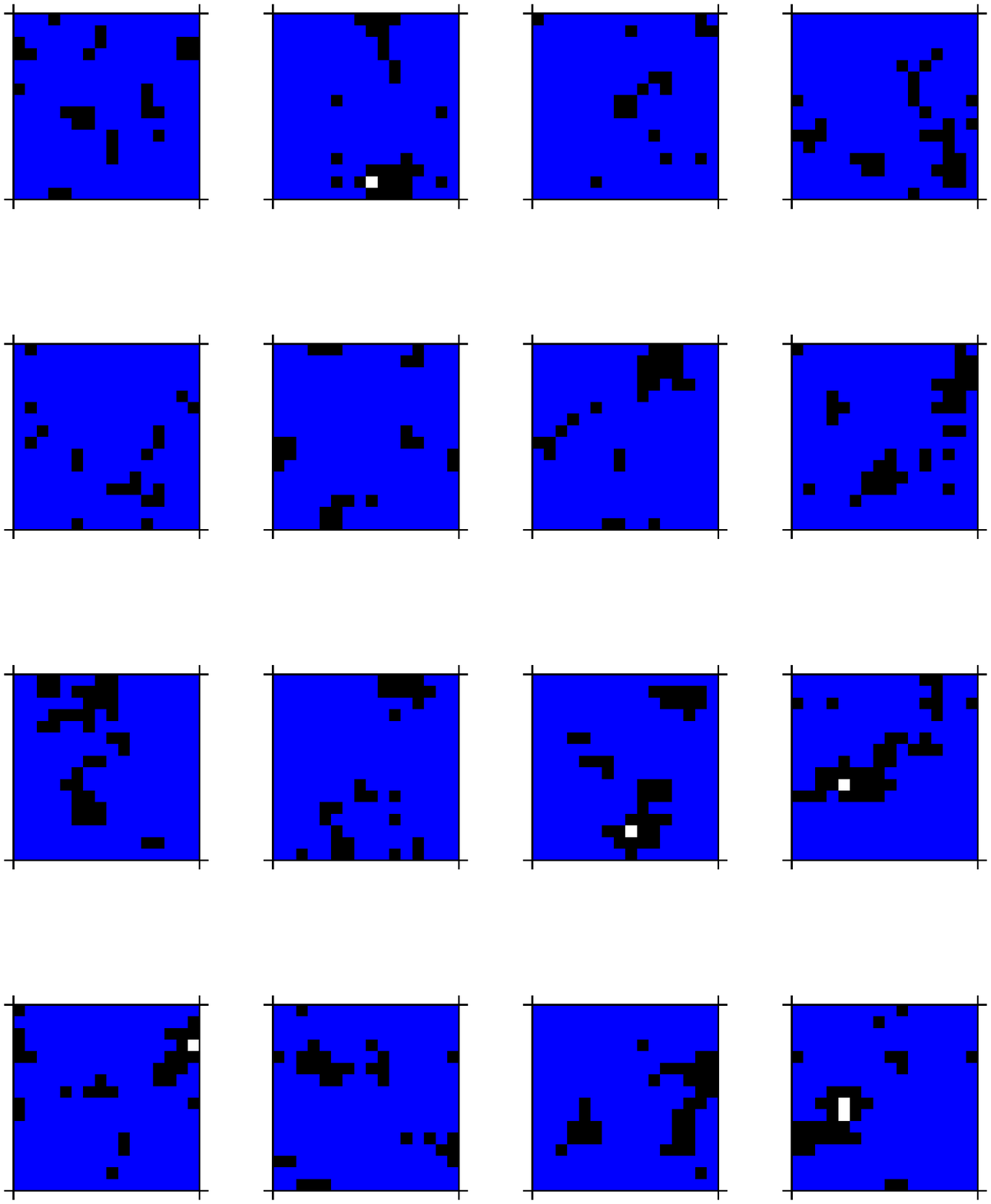}}
 \caption
    {(Color online) 
 Examples of the  percolation pattern of local density for 32 different realizations with 
 disorder strength $\Delta= 15t$. All the clusters are percolated in this case. Clusters of size $L=16$, density $\rho=1.1$, and interaction $U=80t$.}
\label{Percolation_d15}
\end{figure}

\begin{figure}[!htbp]
 \centerline{\includegraphics[trim={0 2cm 0 0},width=0.5\textwidth]{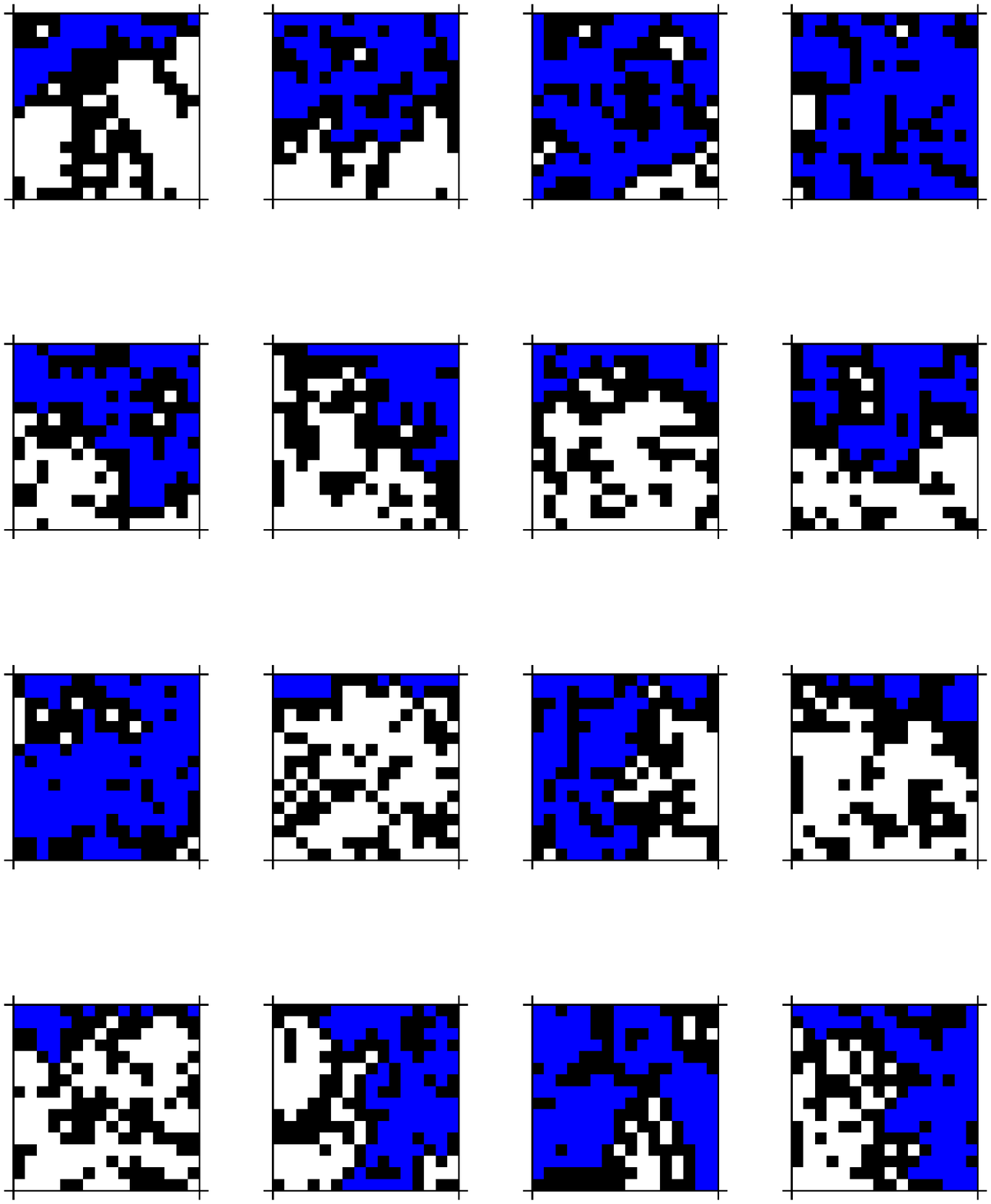}}
  \centerline{\includegraphics[width=0.5\textwidth]{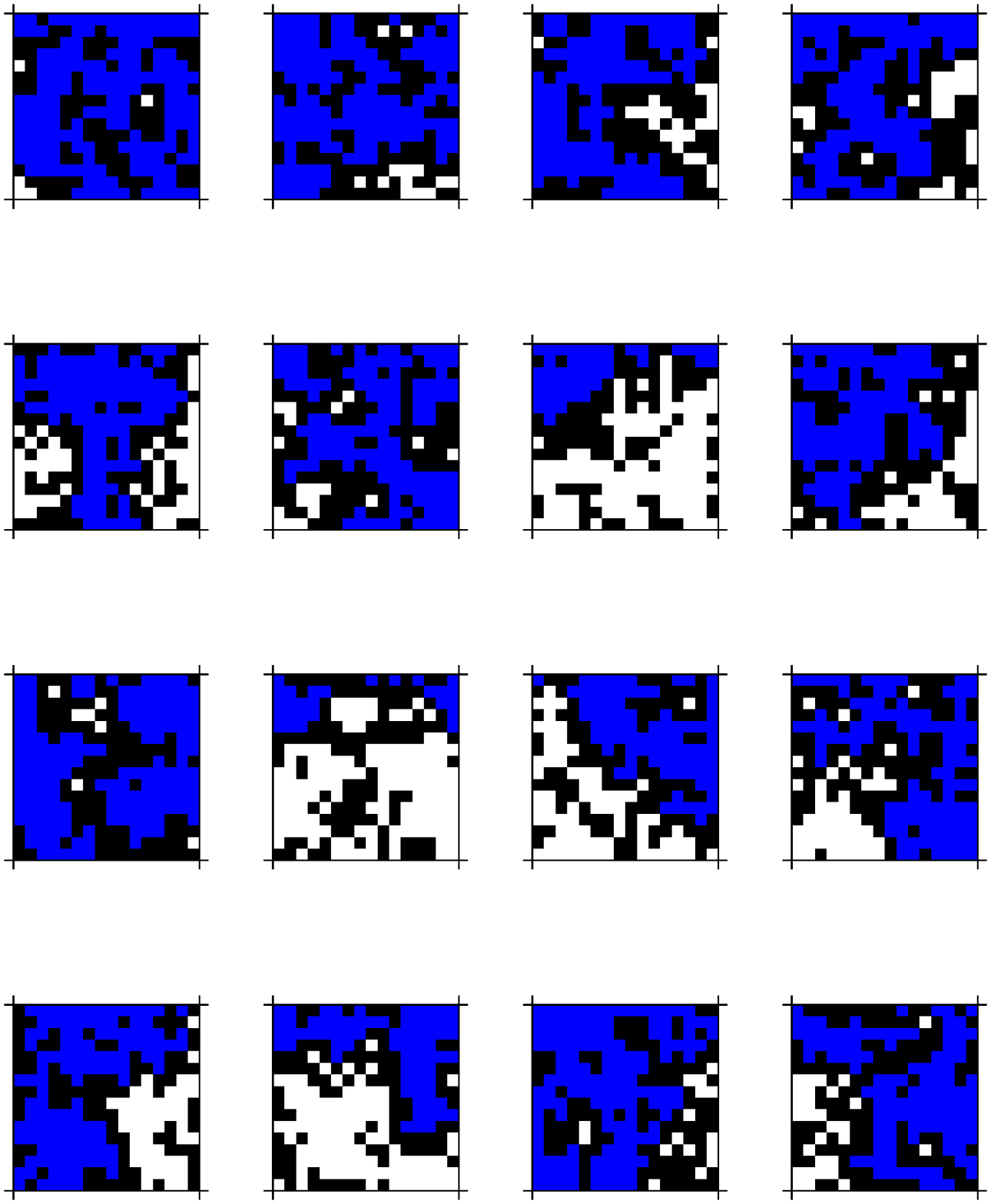}}

 \caption
    {(Color online) 
 Examples of the  percolation pattern of local density for 32 different realizations with 
 disorder strength $\Delta= 25t$. 19 of the clusters are percolated.  Clusters of size $L=16$, density $\rho=1.1$, and interaction $U=80t$.}
\label{Percolation_d25}
\end{figure}

\begin{figure}[!htbp]
 \centerline{\includegraphics[trim={0 2cm 0 0},width=0.5\textwidth]{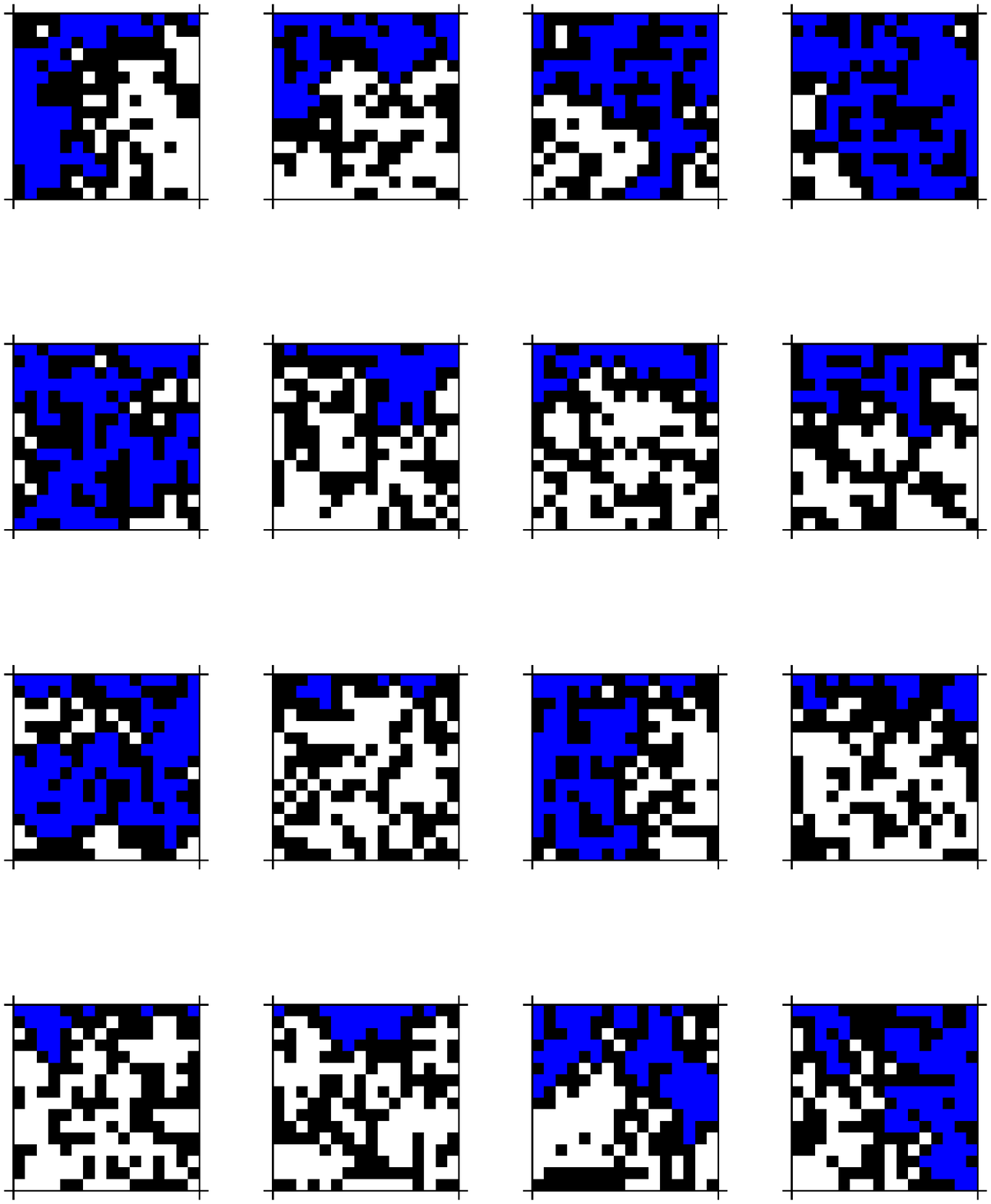}}
 \centerline{\includegraphics[width=0.5\textwidth]{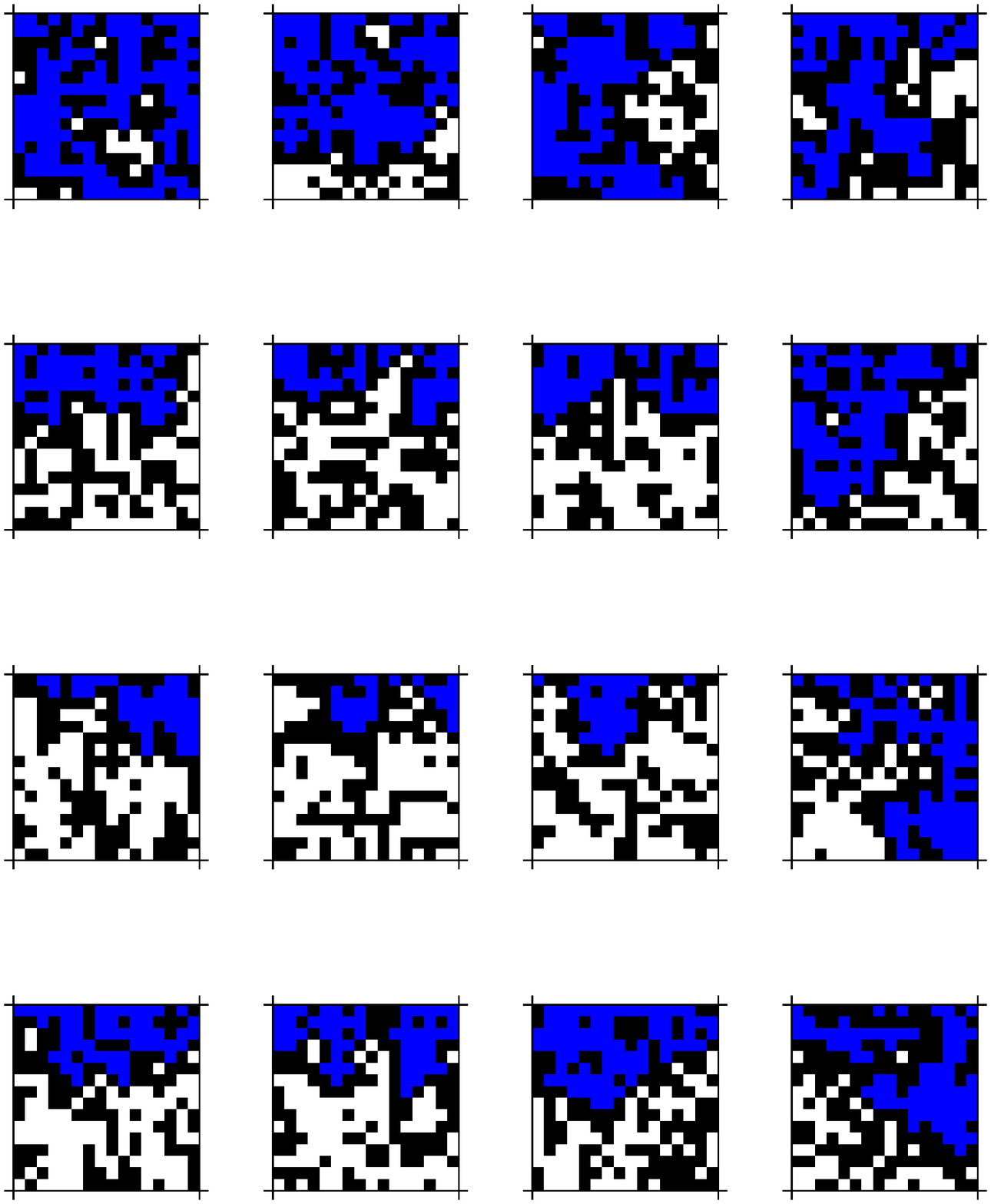}}

 \caption
    {(Color online) 
 Examples of the  percolation pattern of local density for 32 different realizations with 
 disorder strength $\Delta= 30t$. 10 of the clusters are percolated.  Clusters of size $L=16$, density $\rho=1.1$, and interaction $U=80t$.}
\label{Percolation_d30}
\end{figure}

\begin{figure}[!htbp]
 \centerline{\includegraphics[trim={0 2cm 0 0},width=0.5\textwidth]{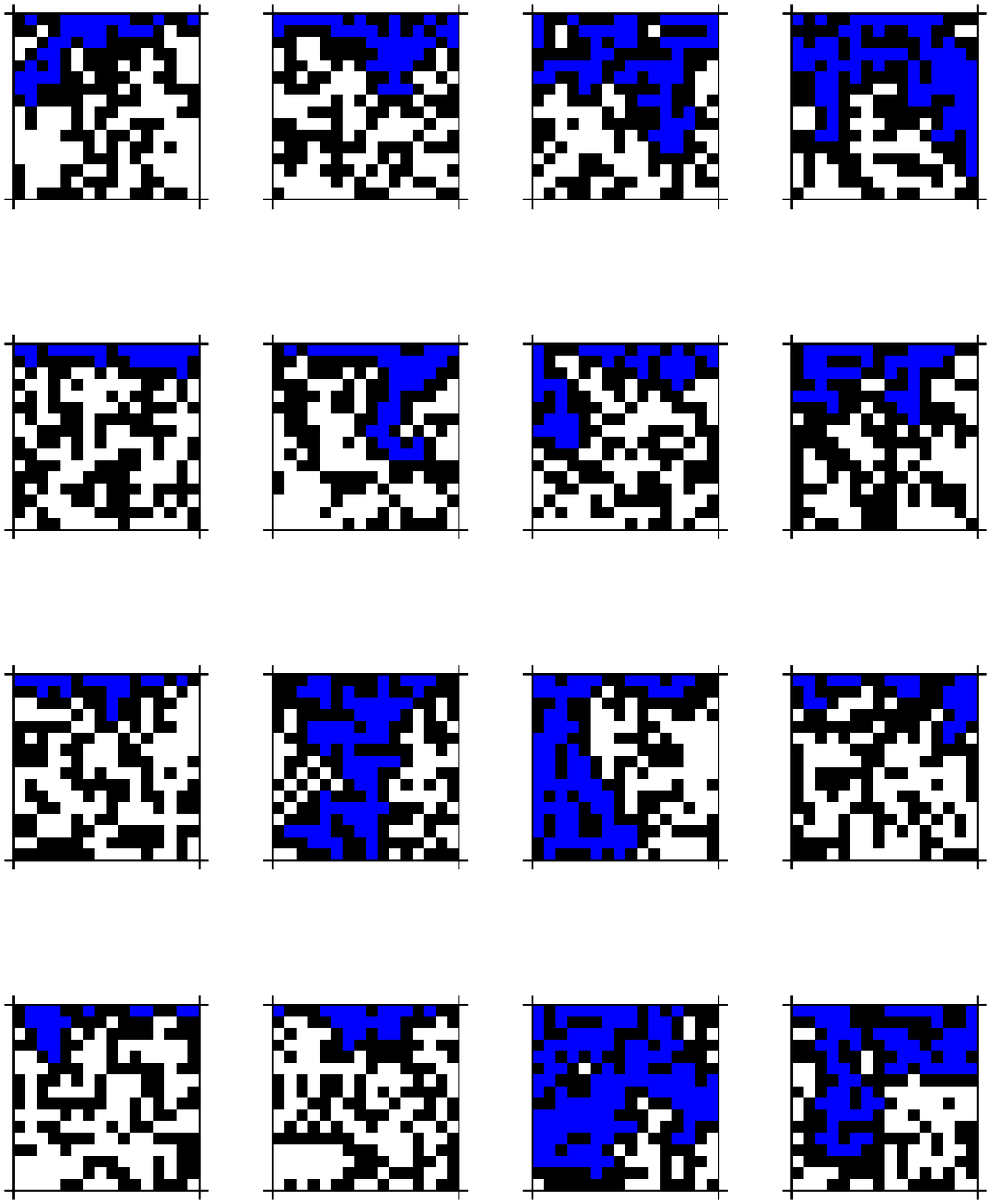}}
  \centerline{\includegraphics[width=0.5\textwidth]{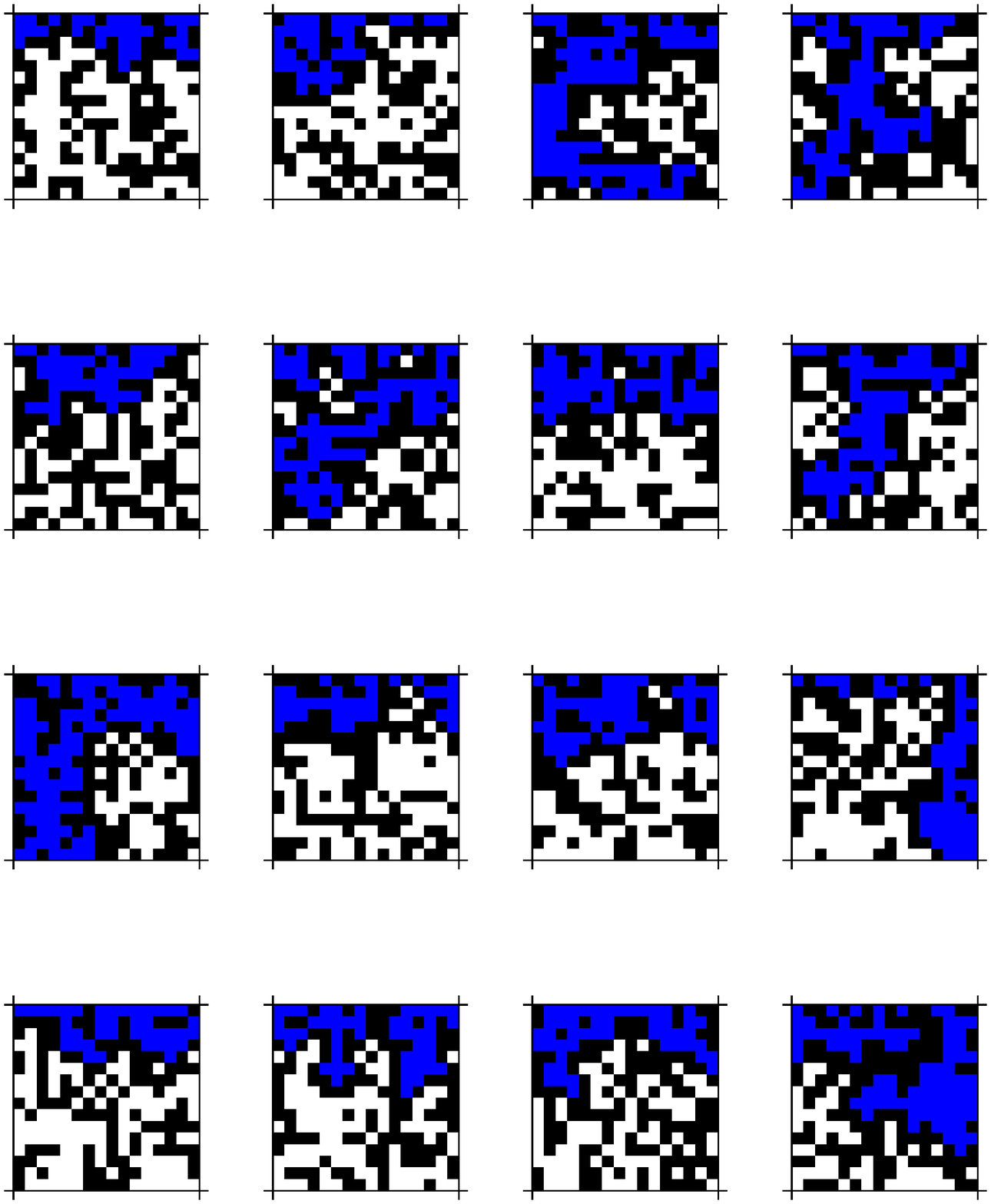}}

 \caption
    {(Color online) 
 Examples of the  percolation pattern of local density for 32 different realizations with 
 disorder strength $\Delta= 35t$. 6 of the clusters are percolated.  Clusters of size $L=16$, density $\rho=1.1$, and interaction $U=80t$.}
\label{Percolation_d35}
\end{figure}



\end{document}